%% file: main.tex
\documentclass[sigconf,screen]{acmart}

\usepackage{xspace}
\usepackage{soul}
\usepackage[ruled,vlined]{algorithm2e}
\usepackage{enumitem}
\usepackage{pifont}

\usepackage{tikz}
\usepackage{xcolor}

\usepackage{booktabs}

\definecolor{golden}{RGB}{230,170,0}

\newcommand{\circnum}[1]{%
\tikz[baseline=(char.base)]{
\node[
shape=circle,
draw=golden,
text=golden,
inner sep=1pt,
line width=0.5pt
] (char) {\small\bfseries #1};}}

\AtBeginDocument{%
  }

\setcopyright{acmlicensed}
\copyrightyear{2026}
\acmYear{2026}
\acmDOI{XXXXXXX.XXXXXXX}
\acmConference[ASE '26]{Proceedings of the 41st IEEE/ACM International Conference on Automated Software Engineering}{October 12--16, 2026}{Munich, Germany}

\acmISBN{978-1-4503-XXXX-X/2026/06}

\newcommand{\toolname}{E-CoDrive\xspace}
\newcommand{\head}[1]{\noindent\textbf{#1.}}

\SetKwComment{Comment}{$\triangleright$\ }{}
\SetCommentSty{itshape}
\newboolean{showcomments}
\setboolean{showcomments}{true}
\ifthenelse{\boolean{showcomments}}
{\newcommand{\nb}[2] {
  \fcolorbox{black}{gray!20}{\bfseries\sffamily\scriptsize#1:}
  {\sf\small$\blacktriangleright$\textit{#2}$\blacktriangleleft$}
}
}
{\newcommand{\nb}[2]{}
}

\begin{document}

\title{\toolname: A Co-Simulation Framework for Testing Energy-Critical Driving Scenarios}

\author{Manfredi Napolitano}
\affiliation{
  \institution{Università degli Studi di Napoli Federico II}
  \city{Naples}
  \country{Italy}
}

\author{Alessandra Somma}
\affiliation{
  \institution{Università degli Studi di Napoli Federico II}
  \city{Naples}
  \country{Italy}
}

\author{Alessio Gambi}
\affiliation{
 \institution{Austrian Institute of Technology}
 \city{Vienna}
 \country{Austria}}
\affiliation{
 \institution{The Italian Institute of AI for Industry}
 \city{Turin}
 \country{Italy}}

\author{Andrea Stocco}
\affiliation{
 \institution{Technical University of Munich}
 \institution{fortiss GmbH}
 \city{Munich}
 \country{Germany}}

\author{Nicola Mazzocca}
\affiliation{
  \institution{Università degli Studi di Napoli Federico II}
  \city{Naples}
  \country{Italy}
}
 
\renewcommand{\shortauthors}{Napolitano et al.}


\begin{abstract}
Autonomous driving research has largely focused on safety while giving limited attention to non-functional aspects such as energy consumption and sustainability.
As Autonomous Electric Vehicles (AEVs) become increasingly common in urban traffic, understanding how complex traffic dynamics influence their energy consumption is paramount to test whether AEVs can complete trips before battery depletion.
To support energy-aware scenario-based testing of AEVs, we present \toolname,
a framework for reproducible closed-loop driving co-simulations that integrates an energy  consumption model, a micro-traffic simulator, and a high-fidelity driving simulator to test AEV software stacks in urban scenarios.
This tool paper describes the architecture of \toolname and 
demonstrates its applicability by testing an Autoware-based AEV stack.
Our evaluation shows that varying traffic conditions produce substantial differences in vehicle energy consumption.
The artifact is publicly available at \url{https://doi.org/10.6084/m9.figshare.32244783},
and a screencast showing the tool is available at \url{https://youtu.be/yX9fWHqCvgc}.
\end{abstract}

\maketitle

\input{sections/01_introduction}
\input{sections/02_framework}

\input{sections/03_example_workflow}

\input{sections/04_preliminary_evaluation}
\input{sections/05_conclusion}

\section{Acknowledgments}

This work was partly funded by MUR (D.D. No. 307, PN RIC 2021–2027, Project ECHO TWIN), co-funded by the European Union (ERDF). Azione 1.1.2\_CUP master: B99H26000290007; Azione 1.1.3b\_CUP master: B92F26000440005, Azione 1.4.3\_CUP master: B89J26003490005, and by the Bavarian Ministry of Economic Affairs, Regional Development and Energy.

\section{Data Availability Statement}

A replication package with our framework, evaluation scripts, and datasets is available at \url{https://doi.org/10.6084/m9.figshare.32244783}. 

\bibliographystyle{ACM-Reference-Format}
\bibliography{bibliography}

\end{document}

%% file: sections/01_introduction.tex
\section{Introduction}\label{sec:introduction}

Research on autonomous driving has made substantial progress in the validation of safety-critical behaviors, perception pipelines, motion-planning logic, and the reliability of learning-enabled components~\cite{survey-lei-ma,lou2022testing,2026-Guo-OJ-ITS,2020-Riccio-EMSE,2020-Humbatova-ICSE}. However, investigating how complex environments and dynamic traffic affect energy consumption remains underexplored, despite its central role in the design of Autonomous Electric Vehicles (AEVs).
Consequently, validating AEVs using existing tools that focus on safety but are oblivious to energy consumption might fail to expose misbehaviors such as idling, repeated acceleration, or detours that lead to premature battery exhaustion.
Studying the interplay among traffic, energy consumption, and recovery requires coordinating traffic generation, autonomous decision-making, and energy-aware measurements in a consistent and reproducible setting.

Unfortunately, existing technologies offer limited support for integrating heterogeneous simulators and for energy-aware scenario-based testing. 
Driving simulators such as CARLA provide high-fidelity vehicle and environment modeling~\cite{dosovitskiy2017}, while SUMO~\cite{krajzewicz2012}, CommonRoad~\cite{althoff2017}, and OpenSBT~\cite{opensbt} support scalable traffic generation and scenario construction. 
Recent work has highlighted that the realism~\cite{2023-Stocco-TSE} and trustworthiness of simulation-based ADS testing critically depend on the coordinated use and fidelity of the underlying simulation environments~\cite{biagiola2023better,sorokin2025simulatorensemblestrustworthyautonomous}. However, these tools are often used in isolation~\cite{zhang2024,HUANG2025104871}. Although they can be integrated~\cite{kato2018} and extended with detailed energy models~\cite{mmpevem2021}, existing solutions primarily focus on data exchange and offer limited support for coordinated execution~\cite{gomes2018cosimulation}. As a result, they may not ensure synchronized execution across traffic, vehicle dynamics, and autonomy modules. This limits their ability to accurately study how variations in traffic induce emergent effects in energy consumption~\cite{survey-lei-ma,lou2022testing}.

To address this gap, we designed \toolname, a novel framework for energy-aware scenario-based testing of AEVs. 
\toolname targets researchers and developers evaluating AEV software stacks under controlled and reproducible traffic conditions. It provides a unified orchestration layer that maintains a shared simulation timeline, coordinates state exchange across heterogeneous components, and manages the end-to-end simulation workflow. This is instrumental in enabling the testing of AEVs by generating and executing energy-aware driving scenarios through synchronized closed-loop co-simulation across traffic, autonomy, and physics, and exposing energy-critical behaviors. 
Our preliminary evaluation shows that, compared with free-flow driving, traffic scenarios executed in \toolname consistently trigger distinct and more variable energy-consumption patterns due to stop-and-go behavior.

By complementing existing research on safety-oriented testing with the analysis of non-functional properties related to energy consumption, this paper makes the following main contributions:

\head{\toolname} A co-simulation framework that integrates traffic generation (SUMO), high-fidelity vehicle and environment simulation (CARLA), and an autonomous driving stack (Autoware Mini) within a unified orchestration layer. 

\head{Scenario Generation} The framework supports the systematic generation of energy-critical driving scenarios, including configurable traffic conditions, vehicle behavior, and energy models.

\head{Reproducible Evaluation} The implementation leverages open-source technologies and enables reproducible execution of scenarios to evaluate how traffic conditions affect energy consumption.

%% file: sections/02_framework.tex
\section{\toolname Architecture}
\label{sec:framework}

\begin{figure}[t]
    \centering
    \includegraphics[width=\linewidth]{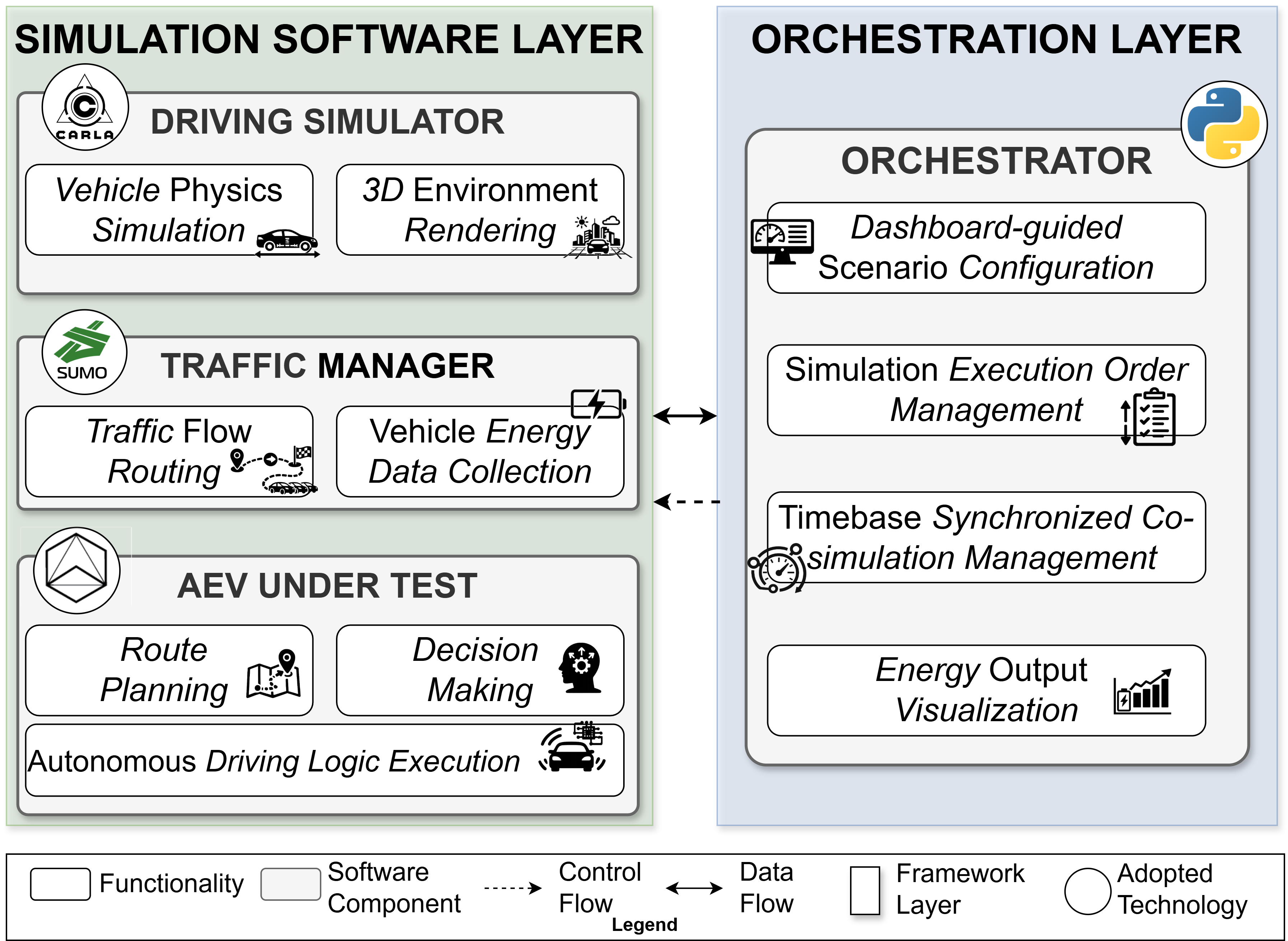}
    \caption{High-level architecture of \toolname.}
    \Description{Block diagram showing the Orchestrator, SUMO, CARLA, and Autoware Mini. The Orchestrator provides dashboard-guided scenario setup, manages execution order, configures synchronized co-simulation, and visualizes energy outputs. SUMO acts as a traffic manager, a routing toolkit, and a source of vehicle energy data. CARLA provides vehicle-physics simulation and 3D environment rendering. Autoware Mini provides the autonomous-driving software for the ego vehicle.}
    \label{fig:architectural_workflow}
\end{figure}

\toolname builds complex simulation scenarios in which AEVs operate under varying traffic conditions. It integrates traffic generation, high-fidelity vehicle and environment simulation, and an autonomous driving stack within a unified orchestration layer. Thanks to \toolname, developers create, execute, and analyze scenarios to study how traffic dynamics influence the AEV's energy demand.

Figure~\ref{fig:architectural_workflow} illustrates the high-level architecture of \toolname. The framework is organized into two main layers. The first is the \textit{simulation software layer}, which encompasses the core simulation tools, including the traffic management system (SUMO), the high-fidelity driving simulator (CARLA), and the AEV under test (Autoware Mini). The second is the \textit{orchestration layer}, which comprises the orchestrator components that coordinate and synchronize interactions among these tools and the components for setting up and visualizing the energy-aware scenarios.

Note that, although the current implementation relies on specific technologies, the framework is modular and configurable, enabling the integration of alternative simulators and autonomous driving stacks. The chosen components reflect their broad adoption and popularity within the autonomous driving community.

\head{Driving Simulator}
CARLA provides 3D environment rendering and vehicle-physics simulation in which both the ego AEV and the surrounding traffic are embedded. Therefore, CARLA serves as the high-fidelity environment in which SUMO-generated traffic and Autoware Mini-issued commands are executed~\cite{dosovitskiy2017}.

\head{Traffic Manager}
SUMO implements the traffic management and route generation. We opted to use it because SUMO offers flexible traffic-demand generation for heterogeneous 
traffic flows and enables the production and collection of vehicle-level energy data when paired with suitable electric-vehicle models~\cite{krajzewicz2012, mmpevem2021}.

\head{Ego Vehicle} 
The AEV under test is controlled by Autoware Mini~\cite{autoware_mini}, which provides the autonomous driving software stack for motion planning and vehicle control. We adopt Autoware Mini because it offers a lightweight, modular, and easily integrable autonomy stack~\cite{kato2018}. In particular, it represents a Python-based re-implementation of the Autoware stack~\cite{autoware-foundation-github}, while still providing realistic deployment characteristics, as it has been deployed on a Lexus RX450h platform at the University of Tartu~\cite{adlvehicle}.

\head{Orchestrator}
It is the central integration component in \toolname.
The orchestrator bridges interactions between users and the underlying simulation layer, hiding from them the complexity of coordinating and configuring executions.
Through a dashboard, the user builds scenarios 
by defining the traffic conditions to be reproduced, including the amount and composition of SUMO-controlled vehicles, the relevant roads in the scenario, and the time window in which vehicles are spawned. Users can also specify the ego's AEV energy model (e.g., MMPEVEM~\cite{mmpevem2021}) and driving task, including the vehicle's static parameters, the initial battery level, and the intended route's starting and destination points. 
After collecting the user configuration, invoking the required routing and scenario-generation steps, and launching the different tools, the orchestrator sets up the synchronized co-simulation. It does so by configuring the interfaces through which SUMO, CARLA, and Autoware Mini advance the simulation according to a common fixed time base. Once the simulation ends, the orchestrator collects the simulation data and visualizes it. Particularly interesting for AEV developers are the visualization of the main energy-related metrics and the driving statistics (e.g., average, minimum, and maximum speed).

%% file: sections/03_example_workflow.tex
\section{E-CoDrive Workflow: An Example Scenario}
\label{sec:exampleworkflow}

\begin{figure}[t]
    \centering
    \includegraphics[width=0.94\linewidth]{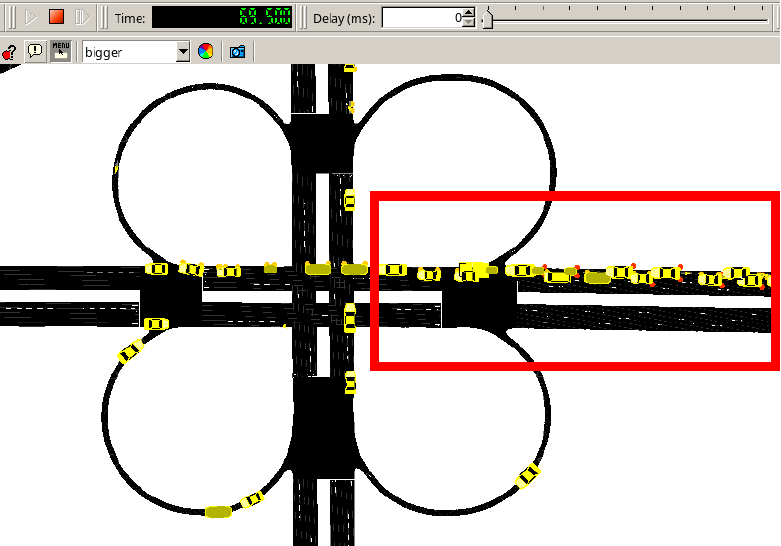}
    \caption{\toolname: Congested-road scenario in SUMO.}
    \label{fig:sumo}
\end{figure}

\begin{figure*}[t]
    \centering
    \includegraphics[width=\textwidth]{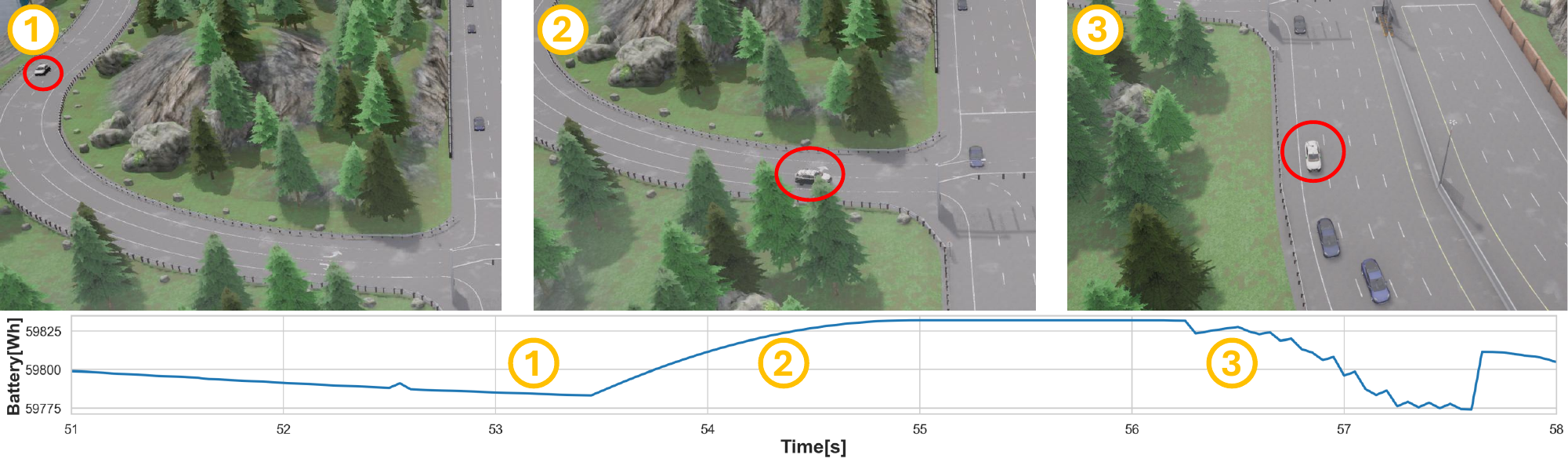}
    \caption{Representative simulated scenario frames highlighting three energy-relevant states of the AEV.}
    \label{fig:energy_scenario}
\end{figure*}

This section illustrates \toolname's workflow using a working example.
First, the user selects an available map and launches CARLA. The current implementation relies on CARLA~v.~\textit{0.9.13} to ensure compatibility with Autoware Mini and offers \textit{Town01}, \textit{Town04}, and \textit{Town05} maps, as these are available for both CARLA and SUMO.

Second, the user defines the traffic surrounding the ego vehicle by specifying start and destination positions and vehicle class\footnote{The vehicle class can be fixed or randomly sampled from the CARLA Vehicle Catalogue available at \url{https://carla.readthedocs.io/en/latest/catalogue_vehicles/}} for all SUMO-controlled vehicles, thereby creating the intended traffic patterns (e.g., a traffic congestion in a selected part of the map, as highlighted by the red frame in Figure~\ref{fig:sumo}). 
This step relies on SUMO route-generation tools, in particular \textit{duarouter}, which generates \texttt{custom\_traffic.rou.xml}, a file containing all traffic routes. The orchestrator then builds \texttt{custom.sumocfg}, the SUMO configuration file used to load the map, routes, and energy output.
The generated scenario is then loaded into SUMO and prepared for co-simulation.

Third, the user configures the energy model of the AEV under test by selecting one of the models, i.e., Energy~\cite{Kurczveil2014} or MMPEVEM~\cite{mmpevem2021}, that are currently available in SUMO.\footnote{\url{https://sumo.dlr.de/docs/Models/Emissions.html}}
These models estimate the energy consumption of electric vehicles as they traverse the map during the simulation.
Based on this choice, the orchestrator generates the static vehicle parameters required by the selected energy model, including battery capacity, vehicle mass, propulsion and recuperation efficiencies, and aerodynamic and rolling coefficients. 
To quickly explore different scenarios related to energy consumption, the user also sets the battery's initial charge level and a threshold below which the battery is considered depleted. This step generates the file \texttt{egovtype.xml}, which SUMO uses to associate the selected energy model with the AEV under test, and the file \texttt{vtypes.json}, which CARLA uses to select the corresponding vehicle models. 

Fourth, the user selects the start and destination positions 
defining the AEV driving task, and launches Autoware Mini with the corresponding vehicle configuration and route. This step uses the \texttt{vtypes.json} file generated previously to link the AEV to the correct vehicle model. 

Finally, the co-simulation starts with SUMO managing the traffic, Autoware Mini driving the AEV under test, and CARLA coordinating the execution using a fixed simulation step of 0.05~s, corresponding to an update frequency of 20~Hz at which all simulators are synchronized before advancing to the next step.
As output, this step produces \texttt{battery.out.xml}, a file that reports instantaneous and cumulative energy consumption, speed, acceleration, and recuperation, including energy recovered during braking events.

\begin{table*}[t]
\centering
\caption{Energy and speed metrics for each traffic scenario and setting.}
\label{tab:battery_metrics}
\resizebox{\textwidth}{!}{%
\begin{tabular}{lrrrrrrrrrrrr}
\toprule
& \multicolumn{4}{c}{Total energy consumed/regen (Wh) $\downarrow$ } & \multicolumn{4}{c}{Actual battery capacity (Wh) $\uparrow$} & \multicolumn{4}{c}{Average speed (m/s) $\uparrow$ } \\
\cmidrule(lr){2-5}\cmidrule(lr){6-9}\cmidrule(lr){10-13}
 & baseline & low & medium & high & baseline & low & medium & high & baseline & low & medium & high \\
\midrule
Town01 & 928/646 & 1594/1207 & 2457/2067 & 2598/2132 & 59721 & 59614 & 59609 & 59587 & 9.77 & 5.73 & 7.30 & 3.69 \\
Town04 & 522/233 & 5585/4911 & 5221/4574 & 4394/3806 & 59711 & 59327 & 59353 & 59413 & 12.81 & 11.48 & 10.61 & 8.59 \\
Town05 & 254/107 & 920/720 & 2056/1766 & 1391/1156 & 59853 & 59801 & 59709 & 59765 & 13.19 & 12.74 & 12.63 & 6.45 \\
\bottomrule
\end{tabular}
}
\end{table*}

%% file: sections/04_preliminary_evaluation.tex
\section{Preliminary Evaluation}
\label{sec:preliminaryeval}

This section presents a preliminary evaluation of our tool, which aims to show that different traffic conditions affect the driving pattern and energy behavior of the AEV under test.
In particular, we consider scenarios that induce congestion near the AEV's destination, thus exposing variations in consumed energy, regenerated energy, and speed. Figure~\ref{fig:energy_scenario} illustrates representative traffic-induced effects on energy consumption at three timestamps, which correspond to free-flow, moderate congestion, and heavy congestion conditions. We highlight the AEV in the figure using a red circle.
In frame {\color{golden}\circnum{1}}, the AEV proceeds without obstacles ahead, and the battery-state profile evolves approximately linearly. In frame {\color{golden}\circnum{2}}, the vehicle approaches the intersection and decelerates, leading to energy recovery through regenerative braking. Finally, in frame {\color{golden}\circnum{3}}, the AEV queues in traffic, and the battery-state profile becomes irregular, potentially indicating an energy-related misbehavior. This example illustrates the relationship between traffic conditions and the energy quantities monitored during simulation.

Building on the pipeline described above, we implemented three application scenarios defined on the CARLA maps adopted by the current implementation. For each map, the AEV route is defined by selecting the two farthest road edges, thereby creating a long mission that we execute under varying traffic conditions.
In each map, we define three traffic levels, denoted as \textit{low}, \textit{medium}, and \textit{high}, corresponding to 10, 25, and 50 vehicles surrounding the AEV under test, respectively. In all cases, the generated traffic is configured to congest the last segment of the AEV's route.
We configured the AEV with an energy model that approximates a Tesla Model 3 AWD profile, using a nominal battery capacity of 75~kWh. Each run starts from a fixed initial charge level of 80\% (i.e., 60~kWh), rather than from a fully charged battery, to preserve plausible headroom for regenerative braking as described by Chandak et al.~\cite{chandak2017}.

We compare the resulting executions against a baseline scenario in which the AEV is the only vehicle present in the network (i.e., free-flow driving). The analysis, therefore, focuses on the variation induced by traffic, examining differences in total energy consumed and regenerated, actual battery capacity, and average vehicle speed relative to the baseline. The energy metrics jointly capture traffic-induced energy exchanges and the resulting net battery depletion, while the average speed indicates whether these variations coincide with degraded driving conditions. Table~\ref{tab:battery_metrics} reports these metrics for each scenario traffic (\textit{low}, \textit{medium}, and \textit{high}), map (\textit{Town01}, \textit{Town04}, and \textit{Town05}), and the baseline run.

Across the three maps, the baseline scenario consistently exhibits the lowest total energy consumption. In general, the gap relative to the baseline increases as traffic density increases, suggesting that estimating energy feasibility of an AEV mission becomes more challenging under congested conditions. A small deviation from this trend is observed for \textit{Town04} and \textit{Town05}, where the measured energy consumption for \textit{high} traffic is slightly lower than that for \textit{medium} traffic. We speculate that larger maps distribute traffic more broadly, thus reducing congestion along the AEV route despite the additional vehicles present on the map.

While no strictly monotonic relationship emerges between consumed and regenerated energy across all configurations, the results consistently show that traffic conditions produce energy behaviors that are substantially different from the baseline scenario. Compared with free-flow driving, traffic scenarios generally exhibit higher energy consumption and regeneration, along with lower average speed, indicating more frequent acceleration and braking phases due to congestion and queuing.

Overall, these results suggest that traffic has a non-negligible impact on the AEV's energy demand. In the most pronounced cases, the increase in battery depletion reaches approximately $137-358$~Wh over routes spanning only $1.3-2.1~km$. Although preliminary, these findings indicate that the interaction between traffic conditions and battery consumption is worth further systematic investigation. In particular, denser traffic appears to induce more frequent acceleration, braking, and speed adaptations, which may substantially affect short-term energy usage and overall route efficiency. These observations motivate future studies on larger and more diverse traffic scenarios to better understand the relationship between traffic dynamics and energy consumption in AEVs.

%% file: sections/05_conclusion.tex
\section{Conclusions and Future Work}\label{sec:conclusion}

This paper presented \toolname, a co-simulation framework for studying how traffic conditions affect the energy behavior of AEVs. By integrating SUMO, CARLA, and Autoware~Mini into a synchronized, guided execution workflow, we support reproducible scenarios, coordinated simulations, and the collection of energy measurements under closed-loop conditions.
The preliminary evaluation shows that traffic dynamics affect the AEV energy profile, battery depletion, and speed relative to the free-flow baseline. This supports traffic-aware energy validation as a relevant non-functional testing objective for AEVs.

Our future plans include extending \toolname to include automated, search-based testing capabilities for battery-critical scenarios. Instead of manually defining traffic conditions, search-based algorithms can systematically explore the scenario space to identify combinations of traffic and energy conditions that lead to problematic behaviors. This is particularly relevant because the number of parameters defining such scenarios is very large. 
Our goal is to automatically generate scenarios in which an AEV that would normally complete its route under free-flow traffic instead experiences delays, stop-and-go behavior, or rerouting conditions that push the battery close to depletion. 
Such scenarios could help identify energy- and traffic-aware failures, including unsafe immobilization, inefficient routing decisions, or behaviors that negatively affect surrounding traffic. In the long term, this could support the development and evaluation of battery-aware planning and fallback strategies.

%% file: bibliography.bib
@inproceedings{krajzewicz2012,
	title        = {Recent Development and Applications of {SUMO} -- Simulation of Urban MObility},
	author       = {Daniel Krajzewicz and Jakob Erdmann and Michael Behrisch and Laura Bieker},
	year         = 2012,
	booktitle    = {International Journal on Advances in Systems and Measurements},
	url          = {https://elib.dlr.de/80483/}
}

@inproceedings{dosovitskiy2017,
	title        = {{CARLA}: {An} Open Urban Driving Simulator},
	author       = {Dosovitskiy, Alexey and Ros, German and Codevilla, Felipe and Lopez, Antonio and Koltun, Vladlen},
	year         = 2017,
	month        = {13--15 Nov},
	booktitle    = {Proceedings of the 1st Annual Conference on Robot Learning},
	publisher    = {PMLR},
	series       = {Proceedings of Machine Learning Research},
	volume       = 78,
	pages        = {1--16},
	url          = {https://proceedings.mlr.press/v78/dosovitskiy17a.html},
	editor       = {Levine, Sergey and Vanhoucke, Vincent and Goldberg, Ken}
}

@inproceedings{althoff2017,
	title        = {CommonRoad: Composable benchmarks for motion planning on roads},
	author       = {Althoff, Matthias and Koschi, Markus and Manzinger, Stefanie},
	year         = 2017,
	booktitle    = {2017 IEEE Intelligent Vehicles Symposium (IV)},
	pages        = {719--726},
	doi          = {10.1109/IVS.2017.7995802}
}

@inproceedings{Kurczveil2014,
	title        = {Implementation of an Energy Model and a Charging Infrastructure in SUMO},
	author       = {Kurczveil, Tam{\'a}s and L{\'o}pez, Pablo {\'A}lvarez and Schnieder, Eckehard},
	year         = 2014,
	booktitle    = {Simulation of Urban Mobility},
	publisher    = {Springer Berlin Heidelberg},
	pages        = {33--43},
	doi          = {10.1007/978-3-662-45079-6_3},
	isbn         = {978-3-662-45079-6}
}

@inproceedings{mmpevem2021,
	title        = {Accurate physics-based modeling of electric vehicle energy consumption in the SUMO traffic microsimulator},
	author       = {Koch, Lucas and Buse, Dominik S. and Wegener, Marius and Schoenberg, Sven and Badalian, Kevin and Dressler-, Falko and Andert, Jakob},
	year         = 2021,
	booktitle    = {2021 IEEE International Intelligent Transportation Systems Conference (ITSC)},
	volume       = {},
	number       = {},
	pages        = {1650--1657},
	doi          = {10.1109/ITSC48978.2021.9564463}
}

@inproceedings{kato2018,
	title        = {Autoware on Board: Enabling Autonomous Vehicles with Embedded Systems},
	author       = {Shinpei Kato and Shota Tokunaga and Yuya Maruyama and Seiya Maeda and Manato Hirabayashi and Yuki Kitsukawa and Abraham Monrroy and Tomohito Ando and Yusuke Fujii and Takuya Azumi},
	year         = 2018,
	booktitle    = {2018 ACM/IEEE 9th International Conference on Cyber-Physical Systems (ICCPS)},
	pages        = {287--296},
	doi          = {10.1109/ICCPS.2018.00035}
}

@inproceedings{lou2022testing,
	title        = {Testing of autonomous driving systems: where are we and where should we go?},
	author       = {Lou, Guannan and Deng, Yao and Zheng, Xi and Zhang, Mengshi and Zhang, Tianyi},
	year         = 2022,
	booktitle    = {Proceedings of the 30th ACM Joint European Software Engineering Conference and Symposium on the Foundations of Software Engineering},
	location     = {Singapore, Singapore},
	publisher    = {Association for Computing Machinery},
	address      = {New York, NY, USA},
	series       = {ESEC/FSE 2022},
	pages        = {31–43},
	doi          = {10.1145/3540250.3549111},
	isbn         = 9781450394130,
	url          = {https://doi.org/10.1145/3540250.3549111},
	numpages     = 13
}

@article{survey-lei-ma,
	title        = {A Survey on Automated Driving System Testing: Landscapes and Trends},
	author       = {Shuncheng Tang and Zhenya Zhang and Yi Zhang and Jixiang Zhou and Yan Guo and Shuang Liu and Shengjian Guo and Yan{-}Fu Li and Lei Ma and Yinxing Xue and Yang Liu},
	year         = 2023,
	month        = jul,
	journal      = {{ACM} Trans. Softw. Eng. Methodol.},
	publisher    = {ACM},
	volume       = 32,
	number       = 5,
	pages        = {124:1--124:62},
	doi          = {10.1145/3579642},
	issn         = {1049-331X},
	url          = {https://doi.org/10.1145/3579642},
	bibsource    = {dblp computer science bibliography, https://dblp.org},
	issue_date   = {September 2023},
	articleno    = 124,
	numpages     = 62
}

@article{zhang2024,
	title        = {Energy-Aware Optimization of Connected and Automated Electric Vehicles Considering Vehicle-Traffic Nexus},
	author       = {Zhang, Ying and Chen, Jinchao and You, Tao and Zhang, Yingjie and Liu, Zhaohua and Du, Chenglie},
	year         = 2024,
	journal      = {IEEE Transactions on Industrial Electronics},
	volume       = 71,
	number       = 1,
	pages        = {282--293},
	doi          = {10.1109/TIE.2023.3245204}
}

@article{HUANG2025104871,
	title        = {Life-cycle carbon emissions of autonomous electric vehicles in varying traffic situations},
	author       = {Kai Huang and Ziheng Zhang and Xiang Wang and Yiheng Tao and Zhiyuan Liu},
	year         = 2025,
	journal      = {Transportation Research Part D: Transport and Environment},
	volume       = 146,
	pages        = 104871,
	doi          = {https://doi.org/10.1016/j.trd.2025.104871},
	issn         = {1361-9209},
	url          = {https://www.sciencedirect.com/science/article/pii/S1361920925002810}
}

@misc{autoware-foundation-github,
	title        = {Autoware Core/Universe},
	author       = {Autoware Foundation},
	year         = {2022--2025},
	journal      = {GitHub repository},
	publisher    = {GitHub},
	howpublished = {\url{https://github.com/autowarefoundation/autoware}}
}

@misc{autoware_mini,
	title        = {Autoware Mini},
	author       = {University of Tartu , Autonomous Driving Lab},
	year         = 2026,
	journal      = {GitHub},
	url          = {https://github.com/UT-ADL/autoware_mini/tree/release/nodes/detection/lidar/cluster}
}

@article{biagiola2023better,
	title        = {Two is better than one: digital siblings to improve autonomous driving testing},
	author       = {Biagiola, Matteo and Stocco, Andrea and Riccio, Vincenzo and Tonella, Paolo},
	year         = 2024,
	month        = may,
	journal      = {Empirical Softw. Engg.},
	publisher    = {Kluwer Academic Publishers},
	address      = {USA},
	volume       = 29,
	number       = 4,
	doi          = {10.1007/s10664-024-10458-4},
	issn         = {1382-3256},
	issue_date   = {Jul 2024},
	numpages     = 33
}

@article{sorokin2025simulatorensemblestrustworthyautonomous,
	title        = {Simulator ensembles for trustworthy autonomous driving systems testing},
	author       = {Sorokin, Lev and Biagiola, Matteo and Stocco, Andrea},
	year         = 2026,
	month        = feb,
	journal      = {Empirical Softw. Engg.},
	publisher    = {Kluwer Academic Publishers},
	address      = {USA},
	volume       = 31,
	number       = 4,
	doi          = {10.1007/s10664-026-10821-7},
	issn         = {1382-3256},
	url          = {https://doi.org/10.1007/s10664-026-10821-7},
	issue_date   = {Apr 2026},
	numpages     = 39
}

@misc{adlvehicle,
	title        = {Vehicle},
	author       = {{Autonomous Driving Lab, University of Tartu}},
	year         = 2023,
	note         = {Accessed: 2026-03-06},
	howpublished = {\url{https://adl.cs.ut.ee/lab/vehicle}}
}

@inproceedings{chandak2017,
	title        = {A review on regenerative braking in electric vehicle},
	author       = {Chandak, Gaurav A. and Bhole, A. A.},
	year         = 2017,
	booktitle    = {2017 Innovations in Power and Advanced Computing Technologies (i-PACT)},
	volume       = {},
	number       = {},
	pages        = {1--5},
	doi          = {10.1109/IPACT.2017.8245098}
}

@article{gomes2018cosimulation,
	title        = {Co-Simulation: A Survey},
	author       = {Gomes, Cl\'{a}udio and Thule, Casper and Broman, David and Larsen, Peter Gorm and Vangheluwe, Hans},
	year         = 2018,
	month        = may,
	journal      = {ACM Comput. Surv.},
	publisher    = {Association for Computing Machinery},
	address      = {New York, NY, USA},
	volume       = 51,
	number       = 3,
	doi          = {10.1145/3179993},
	issn         = {0360-0300},
	url          = {https://doi.org/10.1145/3179993},
	issue_date   = {May 2019},
	articleno    = 49,
	numpages     = 33
}

@article{2020-Riccio-EMSE,
	title        = {{Testing Machine Learning based Systems: A Systematic Mapping}},
	author       = {Vincenzo Riccio and Gunel Jahangirova and Andrea Stocco and Nargiz Humbatova and Michael Weiss and Paolo Tonella},
	year         = 2020,
	journal      = {Empirical Software Engineering},
	publisher    = {Springer},
	doi          = {10.1007/s10664-020-09881-0}
}

@inproceedings{2020-Humbatova-ICSE,
	title        = {{Taxonomy of Real Faults in Deep Learning Systems}},
	author       = {Humbatova, Nargiz and Jahangirova, Gunel and Bavota, Gabriele and Riccio, Vincenzo and Stocco, Andrea and Tonella, Paolo},
	year         = 2020,
	booktitle    = {Proceedings of 42nd International Conference on Software Engineering},
	location     = {Seoul, Republic of Korea},
	publisher    = {ACM},
	series       = {ICSE'20},
	doi          = {10.1145/3377811.3380395},
	isbn         = {978-1-4503-7121-6/20/05},
	numpages     = 11,
	acmid        = 3380395
}

@inproceedings{opensbt,
	title        = {OpenSBT: A Modular Framework for Search-based Testing of Automated Driving Systems},
	author       = {Sorokin, Lev and Munaro, Tiziano and Safin, Damir and Liao, Brian Hsuan-Cheng and Molin, Adam},
	year         = 2024,
	booktitle    = {Proceedings of the 2024 IEEE/ACM 46th International Conference on Software Engineering: Companion Proceedings},
	location     = {Lisbon, Portugal},
	publisher    = {Association for Computing Machinery},
	address      = {New York, NY, USA},
	series       = {ICSE-Companion '24},
	pages        = {94–98},
	doi          = {10.1145/3639478.3640027},
	isbn         = 9798400705021,
	url          = {https://doi.org/10.1145/3639478.3640027},
	numpages     = 5
}

@article{2023-Stocco-TSE,
	title        = {{Mind the Gap! A Study on the Transferability of Virtual Versus Physical-World Testing of Autonomous Driving Systems}},
	author       = {Andrea Stocco and Brian Pulfer and Paolo Tonella},
	year         = 2023,
	month        = {apr},
	journal      = {IEEE Transactions on Software Engineering},
	publisher    = {IEEE Computer Society},
	address      = {Los Alamitos, CA, USA},
	volume       = 49,
	number       = {04},
	doi          = {10.1109/TSE.2022.3202311},
	issn         = {1939-3520}
}

@article{2026-Guo-OJ-ITS,
	title        = {Foundation Models in Autonomous Driving: A Survey on Scenario Generation and Scenario Analysis},
	author       = {Gao, Yuan and Piccinini, Mattia and Zhang, Yuchen and Wang, Dingrui and Moller, Korbinian and Brusnicki, Roberto and Zarrouki, Baha and Gambi, Alessio and Totz, Jan Frederik and Storms, Kai and Peters, Steven and Stocco, Andrea and Alrifaee, Bassam and Pavone, Marco and Betz, Johannes},
	year         = 2026,
	journal      = {IEEE Open Journal of Intelligent Transportation Systems},
	volume       = {},
	number       = {},
	pages        = {1--1},
	doi          = {10.1109/OJITS.2026.3660686}
}
